\documentclass[aps,prb,preprint,superscriptaddress,showpacs,preprintnumbers,amsmath,amssymb]{revtex4-1}

\usepackage{amssymb}
\usepackage{amsmath}
\usepackage[dvips]{graphicx}
\usepackage{epstopdf}
\usepackage{comment}
\usepackage{verbatim}

\begin{document}

\preprint{}

\title{Bichiral structure of feroelectric domain wall driven by flexoelectricity}

\author{P. V. Yudin}
\email{petr.yudin@epfl.ch}
\affiliation{Ceramics Laboratory, Swiss Federal Institute of Technology (EPFL), CH-1015 Lausanne, Switzerland}
\author{A. K. Tagantsev}
\affiliation{Ceramics Laboratory, Swiss Federal Institute of Technology (EPFL), CH-1015 Lausanne, Switzerland}
\author {E. A. Eliseev}
\affiliation{Institute for Problems of Materials Science, National Academy of Sciences of Ukraine, 3, Krjijanovskogo str.,UA-03142 Kiev, Ukraine}
\author {A. N. Morozovska}
\affiliation{Institute of Physics, National Academy of Sciences of Ukraine, 46, pr. Nauki, UA-03028 Kiev, Ukraine}
\author{N. Setter}
\affiliation{Ceramics Laboratory, Swiss Federal Institute of Technology (EPFL), CH-1015 Lausanne, Switzerland}
\date{\today}

\begin{abstract}

The influence of flexoelectric coupling on the internal structure of neutral domain walls in tetragonal phase of perovskite ferroelectrics is studied. The effect is shown to lower the symmetry of 180-degree walls which are oblique with respect to the cubic crystallographic axes, while $\{100\}$ and $\{110\}$ walls stay "untouched". Being of the Ising type in the absence of the flexoelectric interaction, the oblique domain walls acquire a new polarization component with a structure qualitatively different from the classical Bloch-wall structure.
In contrast to the Bloch-type walls, where the polarization vector draws a helix on passing from one domain to the other, in the
flexoeffect-affected wall, the polarization rotates in opposite directions on the two sides of the wall and passes through zero in its center. Since the resulting polarization profile is invariant upon inversion with respect to the wall center it does not brake the wall symmetry in contrast to the classical Bloch-type walls.
The flexoelectric coupling lower the domain wall energy and gives rise to its additional anisotropy that is comparable to that conditioned by the elastic anisotropy.
The atomic order-of-magnitude estimates shows that the new polarization component $P_2$ may be comparable with spontaneous polarization $P_s$, thus suggesting that, in general, the flexoelectric coupling should be mandatory included in  domain wall simulations in ferroelectrics.
Calculations performed for barium titanate yields the maximal value of the $P_2$, which is much smaller than that of the spontaneous polarization.
This smallness is attributed to an anomalously small value of a component of the "strain-polarization"  elecrostictive tensor in this material.
\end{abstract}

\maketitle

\section{Introduction}\label{Introduction}

In the light of miniaturization of electronic devices and achievements of nanotechnology, the question of functionality of domain walls in ferroelectrics is an exciting issue. The current limit for nano-scale engineering is of the order of 10s of nanometers. Having sizes typically of few nanometers, domain walls offer unique properties that are not exhibited in the bulk of a ferroic sample. For example, there is experimental evidence that twin domain walls in nonferroelectric CaTiO$_3$ possess spontaneous polarization \cite{Salje2008}.  Also ferroelectric properties are predicted in antiphase boundaries of otherwise nonferroelectric SrTi0$_3$ \cite{Courtens2001}.  As in the trend of miniaturization the next logical step is utilization of single domain wall as a functional element, the fundamental research of the domain wall's internal structure is highly demanded. If we consider for example 180-degree domain wall, which is the junction between two oppositely poled domains, its simplest profile contains only one polarization component (Ising wall). However it is possible that domain walls with additional polarization components are energetically favorable. A classic example is the Bloch wall, where an additional in-wall-plane polarization component arises, resulting in a helical polarization profile \cite{Marton2012}.
Perovskite crystals represent a class of materials with reach symmetry properties allowing a variety of wall structures. As mechanical and electric properties are strongly coupled in ferroelectrics, the domain wall structure is determined by both electric and elastic properties of a material, and taking into account the latter may radically affect the wall structure  \cite{Courtens2001}.
Up to now in the context of neutral ferroelectric domain walls the description of electromechanical coupling was restricted to  the electrostrictive interaction.
The electrostriction considerably influences the stability of Ising walls and introduces anisotropy of the wall energy \cite{Dvorak}, but this effect does not introduce new features in the wall structure. Recent studies \cite{Morozovska2012} reveal a considerable impact of the generalized flexoelectricity (bilinear coupling between the strain and the order parameter gradient) on the wall structure in ferroics. In this paper we examine the effect of the flexoelectricity on electrically neutral ferroelectric domain walls as a function of wall orientation. We consider perovskite-type ferroelectrics in the tetragonal phase and perform numerical calculations for BaTiO$_3$ (BTO). We show that the flexoelectric effect has no impact on 90-degree walls and on 180-degree walls of the $\{100\}$ and $\{110\}$ orientation. At the same time for oblique 180-degree walls the effect leads to a wall with a new type of structure.

\section{Ginsburg-Landau-Devonshire theory }
\label{sec:Generalized}
We consider perovskite material with $m\overline{3}m$ symmetry of the parent phase. The electric displacement field is defined as $\mathbf{D}=\varepsilon_b \mathbf{E}+\mathbf{P}$,  where $\varepsilon_b$ is the background dielectric permittivity, $\mathbf{E}$ is the vector of electric field, $\mathbf{P}$ is the ferroelectric part of the polarization vector (hereafter we use the term of polarization as a shorthand). The $\mathbf{D}$-field satisfies the Poisson equation:

\begin{equation} \label{Electrostatic condition}
div\mathbf{D}=0
\end{equation}
The Gibbs free energy density expanded to sixth powers of polarization including gradient and flexoelectric terms is written as follows \cite{Morozovska2012}.

 \begin{eqnarray} \label{Free Energy}
G=A_{ij}P_{i}P_{j}+B_{ijkl}P_iP_jP_kP_l+C_{ijklmn}P_iP_jP_kP_lP_mP_n+\frac{1}{2}D_{ijkl}\frac{d P_i}{d x_j}\frac{d P_k}{d x_l}- \\ \nonumber -Q_{ijkl}\sigma_{ij}P_k P_l-\frac{1}{2}s_{ijkl}\sigma_{ij}\sigma_{kl}+\frac{1}{2}F_{ijkl}\left(\sigma_{ij}\frac{d P_k}{d x_l}-P_k\frac{d \sigma_{ij}}{d x_l}\right)
\end{eqnarray}
where $A_{ij}=a_1\delta_{ij}$, $B_{ijkl}=\frac{a_{12}}{2}<\delta_{ij}\delta_{kl}>+(a_{11}-a_{12})g_{ijkl}^{(4)}$
, and $C_{ijklmn}=\frac{a_{123}}{6}<\delta_{ij}\delta_{kl}\delta_{mn}>+(a_{112}-\frac{a_{123}}{2})<\delta_{ij}g_{klmn}^{(4)}>+(a_{111}-a_{112}+\frac{a_{123}}{3})g_{ijklmn}^{(6)}$ are the 2nd, 4th, and 6th order dielectric stiffness tensors. With $<>$ we denote symmetrization with respect to interchange of indices: e.g.  $<\delta_{ij}\delta_{kl}>=\frac{1}{3}(\delta_{ij}\delta_{kl}+\delta_{ik}\delta_{jl}+\delta_{il}\delta_{jk})$,

\begin{equation} \label{Gradient term}
D_{ijkl}=D_{12}\delta_{ij}\delta_{kl}+D_{66}(\delta_{ik}\delta_{jl}+\delta_{il}\delta_{jk})+(D_{11}-D_{12}-2D_{66})g_{ijkl}^{(4)}
\end{equation}
is the tensor controlling the correlation effects.
$\sigma_{ij}$ are the components of mechanical stress.
Hereafter the summation over repeating indices is implied, $\delta_{ij}$ is the invariant Kronecker tensor, $g_{ijkl}^{(4)}$ and $g_{ijklmn}^{(6)}$ are invariant tensors for the cubic symmetry.
In the cubic crystallographic axes the tensors $g_{ijkl}^{(4)}$ and $g_{ijklmn}^{(6)}$  have following structures: $g_{ijkl}^{(4)}=1$ if $i=j=k=l$ and $g_{ijkl}^{(4)}=0$ otherwise; $g_{ijklmn}^{(6)}=1$ if $i=j=k=l=m=n$ and $g_{ijklmn}^{(6)}=0$ otherwise.

Electrostriction tensor $Q_{ijkl}$,compliance tensor $s_{ijkl}$ and flexoelectric tensor $F_{ijkl}$ have same structures as $D_{ijkl}$:

\begin{equation} \label{Electrostriction}
Q_{ijkl}=Q_{12}\delta_{ij}\delta_{kl}+\frac{Q_{66}}{4}(\delta_{ik}\delta_{jl}+\delta_{il}\delta_{jk})+(Q_{11}-Q_{12}-\frac{Q_{66}}{2})g_{ijkl}^{(4)}
\end{equation}

\begin{equation} \label{Elastic Energy}
s_{ijkl}=s_{12}\delta_{ij}\delta_{kl}+\frac{s_{66}}{4}(\delta_{ik}\delta_{jl}+\delta_{il}\delta_{jk})+(s_{11}-s_{12}-\frac{s_{66}}{2})g_{ijkl}^{(4)}
\end{equation}

\begin{equation} \label{Elastic Energy}
F_{ijkl}=F_{12}\delta_{ij}\delta_{kl}+\frac{F_{66}}{2}(\delta_{ik}\delta_{jl}+\delta_{il}\delta_{jk})+(F_{11}-F_{12}-F_{66})g_{ijkl}^{(4)}
\end{equation}
From the Gibbs potential \eqref{Free Energy} one obtains equations of state:

\begin{equation}\label{Euler}
    \frac{\partial G}{\partial P_i}-\frac{d}{dx_j}(\frac{\partial G}{\partial P_{i,j}^{\prime}})=0
\end{equation}
For mechanical stresses we apply conditions of mechanical equilibrium:

\begin{equation} \label{Static condition}
\\\frac{\partial \sigma_{ij}}{\partial x_j}=0 \, \,(i,j=1-3)
\end{equation}
In view of presence of stress gradient in the expression \eqref{Free Energy}, strain is defined via the variational derivation:

\begin{equation}
\label{Constitutive Equation}
{\varepsilon _{ij} =-\partial G \mathord{\left/{\vphantom{\varepsilon _{ij} =-\partial G \partial \sigma _{ij} }}\right.\kern-\nulldelimiterspace} \partial \sigma _{ij} } +\frac{d}{dx_k}({\partial G \mathord{\left/{\vphantom{\partial G \partial \sigma _{ij,k}^{'} }}\right.\kern-\nulldelimiterspace} \partial \sigma _{ij,k}^{'} } )
\end{equation}

\section{Statement of the problem for neutral domain walls}
\label{sec:Generalized}

We consider the material in the tetragonal phase, where the spatially homogeneous solution to the set of equations \eqref{Euler},\eqref{Static condition} for mechanically free sample yields six equivalent domain states $\{P_{s},0,0\}$, $\{-P_{s},0,0\}$, $\{0,P_{s},0\}$, etc. with spontaneous polarization $P_{s}=\sqrt{\frac{\sqrt{a_{11}^2 - 3a_{111}a_1} - a_{11}}{3 a_{111}}}$. 90-degree walls separate domains with the angle of $90^o$ (to within the clapping angle) between the polarization vectors; between oppositely poled domains, 180-degree walls are formed. Below we show that the flexoelectricity does not affect properties of $\{100\}$ and $\{110\}$ walls. Condition of mechanical compatibility allows only one type of orientation for $90^o$ walls, namely $\{110\}$. Thus $90^0$ walls are not affected by flexoelectric coupling. That is why we consider only the 180-degree walls.

\subsection{180-degree walls}
\label{subsec:Generalized}

\begin{figure}
    \includegraphics[width=0.6\textwidth]{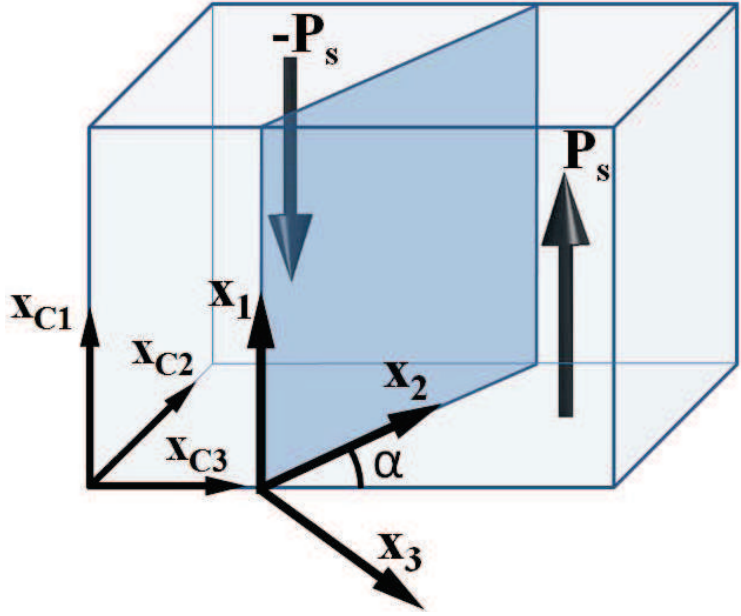}
    \caption{Orientation of the 180-degree domain wall and reference frames used.}
    \label{Coordinates}
\end{figure}

Electrically neutral 180-degree walls are parallel to the $\mathbf{P_s}$ -vector. We characterize the orientation of the wall by the angle $\alpha$ between the wall normal and the $OX_{3C}$ cubic crystallographic direction as shown in Fig. \ref{Coordinates}. Calculations are performed in the reference frame $(0X_1,0X_2,0X_3)$ shown in Fig. \ref{Coordinates}, which is associated with the wall. We consider a one-dimensional (1D) problem with the polarization vector $\textbf{P}$ and mechanical stresses tensor $\sigma_{ij}$ being dependent only on the coordinate $x_3$ normal to the plane of the wall. We neglect the polarization component normal to the wall as suppressed by the strong depolarizing field, so that only $P_1$ and $P_2$ components are allowed. In the new reference frame the Gibbs energy \eqref{Free Energy} reads

\begin{eqnarray} \label{Gibbs_Alp}
G = a_{1} \left(P_{1}^{2} + P_{2}^{2} \right) +a_{11} P_{1}^{4} +a_{22} \left(\alpha \right)P_{2}^{4} +a_{12} P_{1}^{2} P_{2}^{2} +a_{111} P_{1}^{6} + a_{222} \left(\alpha \right)P_{2}^{6}+ \\ \nonumber +a_{112} P_{1}^{4} P_{2}^{2} + a_{122} \left(\alpha \right)P_{2}^{4} P_{1}^{2} +\frac{D_{66} }{2} \left(\frac{\partial P_{1} }{\partial x_{3} } \right)^{2} +\frac{D_{44} \left(\alpha \right)}{2} \left(\frac{\partial P_{2} }{\partial x_{3} } \right)^{2}-\\ \nonumber- Q_{11} \sigma _{1} P_{1}^{2}  - Q_{22} \left(\alpha \right)\sigma _{2} P_{2}^{2} -Q_{66} \sigma _{6} P_{1} P_{2} -Q_{12} \left(\sigma _{1} P_{2}^{2} +\sigma _{2} P_{1}^{2} \right)- \\ \nonumber - \frac{1}{2} \left(s_{11}^{} \sigma _{1}^{2} \, +s_{22}^{} \left(\alpha \right)\sigma _{2}^{2} \, \right)\, -s_{12} \sigma _{1} \sigma _{2} -\frac{1}{2} s_{66} \sigma _{6}^{2} +\frac{1}{2}F_{24} \left(\alpha \right)(\sigma _{2} \frac{\partial P_{2} }{\partial x_{3} }-P_{2} \frac{\partial\sigma _{2} }{\partial x_{3} }). \end{eqnarray}
Here we omitted the terms that are null at $\sigma_3=\sigma_4=\sigma_5=0$, since as we show below in  Subsect. \ref{Mech}, these stress components do not appear in the one-dimensional case. The designations used are:
\begin{equation}
\label{Des Alp}
a_{22} \left(\alpha \right)=a_{11} -\frac{2a_{11} -a_{12} }{4} \sin ^{2} \left(2\alpha \right),                                           \end{equation}

\begin{equation}
\label{Des Q}
Q_{22} \left(\alpha \right)=Q_{11} +\sin ^{2} \left(2\alpha \right)\left(\frac{Q_{66} }{4} -\frac{Q_{11} -Q_{12} }{2} \right),                      \end{equation}

\begin{equation}
\label{Des s}
s_{22} \left(\alpha \right)=s_{11} +\sin ^{2} \left(2\alpha \right)\left(\frac{s_{66} }{4} -\frac{s_{11} -s_{12} }{2} \right),                    \end{equation}

\begin{equation}
\label{Des F}
F_{2223} \left(\alpha \right)\equiv F_{24} \left(\alpha \right)=\frac{\sin \left(4\alpha \right)}{4} \left(F_{66} -F_{11} +F_{12} \right) \equiv F_a\sin \left(4\alpha \right)
\end{equation}

\begin{equation}
\label{Des Alp3}
a_{222} \left(\alpha \right)=a_{111} -\frac{3a_{111} -a_{112} }{4} \sin ^{2} \left(2\alpha \right), \,\, a_{122} \left(\alpha \right)=a_{112} -\frac{2a_{112} -a_{123} }{4} \sin ^{2} \left(2\alpha \right)
\end{equation}

\begin{equation}
\label{Des g}
D_{44} \left(\alpha \right)=D_{66} +\sin ^{2} \left(2\alpha \right)\left(\frac{D_{11} -D_{12} }{2} -D_{66} \right)                                  \end{equation}
The following Voigt's (matrix) notations are used:

$a_{11} \equiv a_{1} $, $a_{1111} \equiv a_{11} $, $6a_{1122} \equiv a_{12} $, $D_{1111} \equiv D_{11} $, $D_{1122} \equiv D_{12} $, $D_{1212} \equiv D_{66} $, $Q_{1111} \equiv Q_{11} $, $Q_{1122} \equiv Q_{12} $, $4Q_{1212} \equiv Q_{66} $, $s_{1111} \equiv s_{11} $, $s_{1122} \equiv s_{12} $, $4s_{1212} \equiv s_{66} $, $F_{1111} \equiv F_{11} $, $F_{1122} \equiv F_{12} $, $2F_{1212} \equiv F_{66} $. \footnotemark[1]

\footnotetext[1]
{Note that unlike common designation of $D_{1212} \equiv D_{1313} \equiv D_{2323} $ as $D_{44} $ in the crystallographic frame, in the laboratory frame with inclined axis one has to distinguish $D_{1313} \equiv D_{55} \equiv D_{1212} \equiv D_{66} $ (independent on the angle $\alpha$) and $D_{2323} \equiv D_{44} $ with the latter depending on the inclination angle $\alpha$.}

From \eqref{Gibbs_Alp} one obtains the equations of state for the polarization components depending only on $x_{3} $ in the form:

\begin{subequations}
\label{Euler Alp}
\begin{eqnarray}
    \label{Euler Alp 1}
    {2a_{1} P_{1} +4a_{11} P_{1}^{3} +2a_{12} P_{2}^{2} P_{1} +6a_{111} P_{1}^{5} +4a_{112} P_{1}^{3} P_{2}^{2} +2a_{122} \left(\alpha \right)P_{1}^{} P_{2}^{4} -} \\ \nonumber {-D_{66} \frac{\partial ^{2} P_{1} }{\partial x_{3}^{2} } -2\left(Q_{11} \sigma _{1} +Q_{12} \sigma _{2} \right)P_{1} -Q_{66}^{} \sigma _{6} P_{2} =0} \\
    \label{Euler Alp 2}
    {2a_{1} P_{2} +4a_{22} \left(\alpha \right)P_{2}^{3} +2a_{12} P_{1}^{2} P_{2} +6a_{222} \left(\alpha \right)P_{2}^{5} +2a_{112} P_{1}^{4} P_{2}^{} +4a_{122} \left(\alpha \right)P_{1}^{2} P_{2}^{3} -} \\ \nonumber {-D_{44} \left(\alpha \right)\frac{\partial ^{2} P_{2} }{\partial x_{3}^{2} } -2\left(Q_{22} \left(\alpha \right)\sigma _{2} +Q_{12}^{} \sigma _{1} \right)P_{2} -Q_{66}^{} \sigma _{6} P_{1} -F_{24} \left(\alpha \right)\frac{\partial \sigma _{2} }{\partial x_{3} } =0}
\end{eqnarray}
\end{subequations}
The boundary conditions for the polarization far from the wall are
\begin{equation}
\label{Pol Bound Cond}
P_{1} \left(x_{3} \to -\infty \right)=-P_{S} ,\quad P_{1} \left(x_{3} \to \infty \right)=P_{S} ,\quad P_{2} \left(x_{3} \to \pm \infty \right)=0.
      \end{equation}

\subsection{Elimination of mechanical variables}
\label{Mech}
We consider the bulk of domains to be mechanically free:
\begin{equation}
\label{Sigma Bound Cond}
 \sigma _{ij} (x_{3} \to \pm \infty )=0. \, i,j=1,2,3.
\end{equation}
This implies, using  Eqs. \eqref{Free Energy} and \eqref{Constitutive Equation}, the boundary conditions for the strain components:
\begin{equation} \label{Spont Strains}
\varepsilon_{11}^{} =Q_{11} P_{S}^{2}; \,\, \varepsilon_{22}^{} =Q_{12} P_{S}^{2}; \,\,\varepsilon_{12}^{} =0.
\end{equation}
For our 1D problem, the condition of  mechanical equilibrium \eqref{Static condition} reads  ${\partial \sigma _{3}  \mathord{\left/{\vphantom{\partial \sigma _{3}  \partial x_{3} }}\right.\kern-\nulldelimiterspace} \partial x_{3} } =0$, ${\partial \sigma _{4}  \mathord{\left/{\vphantom{\partial \sigma _{4}  \partial x_{3} }}\right.\kern-\nulldelimiterspace} \partial x_{3} } =0$, ${\partial \sigma _{5}  \mathord{\left/{\vphantom{\partial \sigma _{5}  \partial x_{3} }}\right.\kern-\nulldelimiterspace} \partial x_{3} } =0$.
In view of \eqref{Sigma Bound Cond}, this condition requires that $\sigma _{3} =\sigma _{4} =\sigma _{5} =0$ everywhere.
The 1D character of the problem, also enables us to rewrite, the Saint-Venant compatibility relationships
\begin{equation}
\label{Saint Venant}
e_{ikl}^{} e_{jmn}^{} \left({\partial ^{2} \varepsilon _{ln}^{}  \mathord{\left/{\vphantom{\partial ^{2} \varepsilon _{ln}^{}  \partial x_{k} \partial x_{m} }}\right.\kern-\nulldelimiterspace} \partial x_{k} \partial x_{m} } \right)=0,
\end{equation}
($e_{ijk}$ is the Levi-Civita symbol) as:

 \begin{equation}
\label{Saint Venant x3}
 {d^{2} \varepsilon _{11}  \mathord{\left/{\vphantom{d^{2} \varepsilon _{11}  dx_{3}^{2} }}\right.\kern-\nulldelimiterspace} dx_{3}^{2} } ={d^{2} \varepsilon _{12}  \mathord{\left/{\vphantom{d^{2} \varepsilon _{12}  dx_{3}^{2} }}\right.\kern-\nulldelimiterspace} dx_{3}^{2} } ={d^{2} \varepsilon _{22}  \mathord{\left/{\vphantom{d^{2} \varepsilon _{22}  dx_{3}^{2} }}\right.\kern-\nulldelimiterspace} dx_{3}^{2} } =0.
 \end{equation}
The solution to Eq. \eqref{Saint Venant x3} with boundary conditions \eqref{Sigma Bound Cond} is:

\begin{equation} \label{Spont Strains}
\varepsilon_{1}(x_3) =Q_{11} P_{S}^{2}; \,\, \varepsilon_{2}(x_3)=Q_{12} P_{S}^{2}; \,\,\varepsilon_{6}(x_3) =0
\end{equation}
Note that it is the only possible one-dimensional solution for the elastic problem.
The applicability of this solution to a stress-free finite sample is equivalent to the applicability of a one-dimensional model to a parallel plate capacitor. By applying this we neglect the fringe elastic fields at the contact of the domain wall with the surface, which is permissible  when the dimensions of the sample are much larger than the thickness of the domain wall. Note that same "partially clamped" elastic conditions are usually applied for the description of mechanical stresses in a thin ferroelectric film on a substrate \cite{Pertsev98}. Under this ansatz we solve the system of equations \eqref{Constitutive Equation} and obtain expressions for the nonzero elastic stress components in the form:

\begin{subequations}
\label{Sigma Through P}
\begin{eqnarray}
\label{Sigma1}
\sigma _{11} \equiv \sigma _{1} =\frac{\left(\begin{array}{l} {-F_{24} \left(\alpha \right)s_{12} \left({\partial P_{2}  \mathord{\left/{\vphantom{\partial P_{2}  \partial x_{3} }}\right.\kern-\nulldelimiterspace} \partial x_{3} } \right)+\left(P_{S}^{2} -P_{1}^{2} \right)\left(Q_{11} s_{22} \left(\alpha \right)-Q_{12} s_{12} \right)} \\ {+P_{2}^{2} \left(Q_{22} \left(\alpha \right)s_{12} -Q_{12} s_{22} \left(\alpha \right)\right)} \end{array}\right)}{s_{22} \left(\alpha \right)s_{11} -s_{12}^{2} }\\
\label{Sigma2}
\sigma _{22} \equiv \sigma _{2} =\frac{\left(\begin{array}{l} {F_{24} \left(\alpha \right)s_{11} \left({\partial P_{2}  \mathord{\left/{\vphantom{\partial P_{2}  \partial x_{3} }}\right.\kern-\nulldelimiterspace} \partial x_{3} } \right)+\left(Q_{12} s_{11} -Q_{11} s_{12} \right)\left(P_{S}^{2} -P_{1}^{2} \right)} \\ {+P_{2}^{2} \left(-s_{11} Q_{22} \left(\alpha \right)+Q_{12} s_{12} \right)} \end{array}\right)}{s_{22} \left(\alpha \right)s_{11} -s_{12}^{2} }\\
\label{Sigma6}
\sigma _{12} \equiv \sigma _{6} =-\frac{Q_{66} }{s_{66} } P_{1} \left(x_{3} \right)P_{2} \left(x_{3} \right)
\end{eqnarray}
\end{subequations}

Eqs. \eqref{Sigma Through P} and \eqref{Euler Alp} form a full set of equations to define the polarization profile.

\section{Analysis of equations and estimates}
\label{Estimations}
First, as a benchmark, let us consider the set \eqref{Euler Alp} and \eqref{Sigma Through P} without the flexoelectric effect (with $F_{24}(\alpha)$ set to zero). One can check that at $F_{24}(\alpha)=0$ the set \eqref{Euler Alp} and \eqref{Sigma Through P} has single-component solution, as Eq. \eqref{Euler Alp 2}  can be satisfied with $P_2=0$. This Ising solution may be either stable or unstable \cite{TagantsevBook}; in the latter case Bloch wall profile is observed with $P_2$ being an even function of $x_3$.  We consider the first case where the Ising profile is stable if the flexoelectric effect is neglected.

Since $F_{24}(\alpha )$ is proportional to $\sin(4\alpha)$ \eqref{Des F}, the flexoelectric effect does not reveal itself for $\{100\}$ ($\alpha=0$) and $\{110\}$ ($\alpha =\pi/4$) wall orientations. Hence, these walls remain Ising with the flexoelectric effect "switched on". For all the other wall orientations (hereafter termed as oblique walls), the flexoelectric coupling inevitably leads to the appearance of an additional polarization component. Indeed, substitution of \eqref{Sigma Through P} into Eq. \eqref{Euler Alp 2} introduces an additional coupling between the polarization components.
As a result the one-component solution for the polarization profile (with $P_2=0$) is no longer available.

It is instructive to note that $\{110\}$ - oriented $90^o$ domain walls do not "feel" the flexoelectric coupling by the same reason.
 The statement of the problem for $90^o$ walls is the same, but with boundary conditions

 \begin{equation}
\label{Pol Bound Cond 90}
P_{1} \left(x_{3} \to \pm \infty \right)=0 ,\quad P_{2} \left(x_{3} \to -\infty \right)=-P_{s}/\sqrt{2} ,\quad P_{2} \left(x_{3} \to \infty \right)=P_{s}/\sqrt{2}
\end{equation}
instead of \eqref{Pol Bound Cond} and with $P_3(x_3)=P_s/\sqrt{2}$. One can carry out the same analysis as we have done for $180^o$-walls and check that flexoelectricity produce no nonzero terms in the equations of state for the polarization vector. Thus  the flexoelectric effect does not affect either $90^o$-walls or $\{100\}$ and $\{110\}$-oriented $180^o$ walls.
In contrast, oblique $180^o$ walls inevitably acquire the new polarization component $P_2$.

\subsection{Estimation of magnitude of flexoelectric-effect-induced $P_2$ component}
\label{subsec:Generalized}

To roughly estimate the magnitude of the $P_2$ component induced by the flexoelectric effect, we consider the linearized Eq. \eqref{Euler Alp 2}, where we use for $P_1$ the profile of Ising wall, not perturbed by flexoelectric coupling. This approximation is fully valid at the vicinity of $\{100\}$ and $\{110\}$ wall orientations, where flexoelectric coupling is small.  Neglecting $P_{2} \left(x_{3} \right)$ with respect to $P_{1} \left(x_{3} \right)$ in Eq.\eqref{Euler Alp 1} we obtain the standard wall profile \cite{Bulaevskii}

\begin{subequations}
\label{P1 Approxim}
\begin{eqnarray}
\label{2-4-6 P1 Apploxim}
P_{1}^{2} = P_{S}^{2} \sinh ^{2} \left({x_{3}  \mathord{\left/{\vphantom{x_{3}  R_{c} }}\right.\kern-\nulldelimiterspace} R_{c} } \right)\left(\cosh ^{2} \left({x_{3}  \mathord{\left/{\vphantom{x_{3}  R_{c} }}\right.\kern-\nulldelimiterspace} R_{c} } \right)+A\right)^{-1}        \\
\label{Rc}
R_{c} =\sqrt{\frac{D_{66} }{P_{S}^{2} \left(2\tilde{a}_{11} +6a_{111} P_{S}^{2} \right)} } \\
\label{A}
A=\frac{2a_{111} P_{S}^{2} }{2\tilde{a}_{11} +4a_{111} P_{S}^{2} }\\
\label{a11sVolnoj}
\tilde{a}_{11} \equiv a_{11} +\frac{Q_{12} \left(Q_{12} s_{11} -Q_{11} s_{12} \right)+Q_{11}^{} \left(Q_{11} s_{22} \left(\alpha \right)-Q_{12} s_{12} \right)}{2s_{22} \left(\alpha \right)s_{11} -s_{12}^{2} }
\end{eqnarray}
\end{subequations}

Linearizing the set of Eqs. \eqref{Sigma Through P} and  \eqref{Euler Alp 2} with respect to $P_2$ we obtain:

\begin{subequations}
\label{Euler2 Linear}
\begin{eqnarray}
\label{Euler2 Linear Main}
(2\tilde{a}_{1} +2\tilde{a}_{12} P_{1}^{2} +2a_{112} P_{1}^{4} )P_{2}^{} -\tilde{D}_{44} \left(\alpha \right)\frac{\partial ^{2} P_{2} }{\partial x_{3}^{2} } =f\left(\alpha \right)\frac{\partial (P_{s}^{2} -P_{1}^{2} )}{\partial x_{3} }\\
\label{Euler2 a1}
\tilde{a}_{1} =a_{1} +\Theta P_{s}^{2} \\
\label{Euler2 a12}
 \tilde{a}_{12} =a_{12} +\frac{Q_{66} }{2s_{66} } -\Theta \\
\label{Teta}
\Theta =\frac{s_{12}^{} \left(Q_{12}^{2} -Q_{11} Q_{22} \left(\alpha \right)\right)-Q_{12} \left(Q_{22} s_{11} -Q_{11} s_{22} \left(\alpha \right)\right)}{2s_{22} \left(\alpha \right)s_{11} -s_{12}^{2} }\\
\label{D Renorm}
\tilde{D}_{44} \left(\alpha \right)\equiv D_{44} \left(\alpha \right)+\frac{F_{24} \left(\alpha \right)^{2} s_{11} }{s_{22} \left(\alpha \right)s_{11} -s_{12}^{2} }\\
\label{f Grand}
f\left(\alpha \right)=\frac{\left(Q_{12} s_{11} -Q_{11} s_{12} \right)}{s_{22} \left(\alpha \right)s_{11} -s_{12}^{2} } F_{24} \left(\alpha \right)
\end{eqnarray}
\end{subequations}
with $P_1$ coming from \eqref{P1 Approxim}.

$P_1$ being an odd function, the symmetry of Eq. \eqref{Euler2 Linear} allows an odd solution for $P_2(x_3)$ component. As we confirm by the numerical calculations below, the odd solution is stable, meaning in particular $P_2=0$; $dP_2/dx_3\neq0$ at $x_3=0$.  Hence, in the vicinity of $x_3=0$, the main contribution to the Landau energy is due to the gradient term, which allows us to derive an approximate solution by neglecting the term linear with respect to $P_2$.
We also set $A\to0$ in Eq.\eqref{Approxim Sol 1} for the simplicity.
On simplifying thus, in the vicinity of $x_3=0$  Eq. \eqref{Euler2 Linear} transforms into

\begin{equation}
\label{Euler2 Approxim 1}
\tilde{D}_{44} \left(\alpha \right)\frac{\partial ^{2} P_{2} }{\partial x_{3}^{2} } =f\left(\alpha \right)\frac{\partial (P_{1}^{2} -P_{s}^{2} )}{\partial x_{3} }.
\end{equation}
The first integral of Eq. \eqref{Euler2 Approxim 1} is

\begin{equation}
\label{Euler2 Approxim 2}
\tilde{D}_{44} \left(\alpha \right)\frac{\partial P_{2} }{\partial x_{3}^{} } =f\left(\alpha \right)\left(\left(P_{1}^{2} -P_{S}^{2} \right)+C_{0} \right).
\end{equation}
We set $C_{0} =0$ to prevent the linear increase of $P_2$ at $x_{3} \to \pm \infty $. Taking into account that $P_2(0)=0$, we obtain the solution in the form:

\begin{equation}
\label{Approxim Sol 1}
P_{2} \left(x_{3} \right)=\frac{f\left(\alpha \right)}{\tilde{D}_{44} \left(\alpha \right)} \int _{0}^{x_{3} }\left(P_{1}^{2} -P_{S}^{2} \right)dx \approx -\frac{f\left(\alpha \right)}{\tilde{D}_{44} \left(\alpha \right)} P_{S}^{2} R_{c} \tanh \left(\frac{x_{3} }{R_{c} } \right)           \end{equation}
The solution \eqref{Approxim Sol 1} does not satisfy the boundary conditions $P_{2} \left(x_{3} \to \pm \infty \right)=0$. It means that the other term that we do not take into account in \eqref{Euler2 Approxim 1} is responsible for the decay of $P_2$ in the domains. However, the value $P_{m} =P_{2} \left(x_{3} =\infty \right)$ from \eqref{Approxim Sol 1} may be used for the estimation of the amplitude of $P_2$. Reverting to the initial designations we obtain $P_m$ in the form:

\begin{equation}
\label{Approxim Amp F}
P_{m} =\frac{F_{24} \left(\alpha \right)}{D_{44} \left(\alpha \right)\left(s_{22} \left(\alpha \right)-\frac{s_{12}^{2} }{s_{11} } \right)+F_{24} \left(\alpha \right)^{2} } \left(Q_{12} -Q_{11} \frac{s_{12} }{s_{11} } \right)P_{S}^{2} R_{c}
\end{equation}
We rewrite it in dimensionless form:

\begin{subequations}
\label{Estim 1}
\begin{eqnarray}
\label{Estim 1 Main}
\frac{P_{m0}}{P_s} =\frac{F_{0}}{\sqrt{D_{66}s_{11} }} \frac{Q_{11}}{\sqrt{2\tilde{a}_{11}s_{11} }}\left(\frac{Q_{12}}{Q_{11}}-\frac{s_{12}}{s_{11}}\right)\cdot\Gamma\left(\alpha \right)\\
\label{Gamma}
\Gamma\left(\alpha \right)=\frac{\sin(4\alpha)D_{66}/D_{44}\left(\alpha \right)}{\left( 1+\frac{F_{24}\left(\alpha \right)^2}{D_{44}\left(\alpha \right)\left(s_{22}\left(\alpha \right)-\frac{s_{12}^2}{s_{11}^2}\right)}\right)\left(1-\frac{s_{12}^2}{s_{11}^2}+\sin^2\left(2\alpha \right)\left(\frac{s_{66}}{4s_{11}}+\frac{s_{12}}{2s_{11}}-\frac{1}{2} \right)\right)}
\end{eqnarray}
\end{subequations}

Let us analyze expression \eqref{Estim 1}. The factor $\frac{F_{0}}{\sqrt{D_{66}s_{11} }}$ is of the order of unity according to atomic estimates.
The applicability of atomic estimates for the evaluation of the electrostrictive tensor in perovskite ferroelectrics is supported by experimental evidence of rather strong flexoelectric coupling in these materials \cite{Axe}.
The factor $\frac{Q_{11}}{\sqrt{\tilde{a}_{11}s_{11} }}$, which is responsible for the sound velocity change near the ferroelectric phase transition is also of the order of unity in ferroelectrics with strong electromechanical coupling \cite{Levanyuk}. Landau theory does not require that the dimensionless factors $\frac{Q_{12}}{Q_{11}}-\frac{s_{12}}{s_{11}}$ and $\Gamma\left(\alpha \right)$ are small compared to unity, so that for rough estimation they may be taken as unity. Thus from formula \eqref{Estim 1} the flexoelectric - driven component $P_2$ is expected to be of the same order than $P_s$.

It is instructive to specify the above estimates for BTO crystal, for which, further in the paper, we will present numerical simulations for the polarization profile of the domain boundary.
For BTO the factor $\frac{Q_{12}}{Q_{11}}-\frac{s_{12}}{s_{11}}$ appears to about $1/15$.
The smallness of this factor might be considered as purely accidental in view of seemingly different physical phenomena behind the elasticity and electrostriction.
However, this smallness is of purely electromechanical nature.
To see this we rewrite the factor $Q_{12}/Q_{11}-s_{12}/s_{11}$ in terms of the "strain-polarization" electrostriction tensor $q_{ij}$ linked  with $Q_{ij}$  by the relationship $Q_{ij}=s_{jl}q_{il}$.
The tensor $q_{ij}$ can be considered as a primary material parameter since it is  directly controlled by the lattice mechanics of the crystal.
Being interested in perovskites where the Poisson ratio $-s_{12}/s_{11}$ is typically about $0.3$, we readily find $Q_{12}/Q_{11}-s_{12}/s_{11} \simeq q_{12}/q_{11}(0.5/(1-0.6(q_{12}/q_{11}))$.
Thus we see the smallness of the factor $Q_{12}/Q_{11}-s_{12}/s_{11}$ is conditioned by the fact that, in metal-oxide ferroelectric perovskites, the ratio $q_{12}/q_{11}$ is typically very small compared to unity. In particular, for $SrTiO_3$ $q_{12}/q_{11}=-0.086$ \cite{PertsevST}, in $Pb(ZrO_3)_{1-x}(TiO_3)_x$ (PZT) $|q_{12}/q_{11}|<0.05$ for $x \in(0.6,1)$ ($q_{11},\,\,q_{12}$ recalculated using Refs. \cite{Haun}, \cite{PertsevS} ) . However there are  materials where $q_{12}$ is not small with respect to $q_{11}$. For example in $KNbO_3$ $q_{12}/q_{11}=-0.37$, in tetragonal PZT near the morphotropic boudary $q_{12}/q_{11}\sim-0.5$ (\cite{Haun}, \cite{PertsevS}). There is no data available on the flexoelectric tensor coefficients in the latter materials, but one can expect there relatively large domain-wall energy anisotropy and flexoelectric-effect-induced polarization components due to the elevated $q_{12}/q_{11}$ ratio.

For BTO the estimate for the amplitude of second polarization components is also affected by the exceptionally high anisotropy of the correlation energy ($D_{11}/D_{66}\simeq 25$).
This occurs via the factor $\frac{D_{66}}{D_{44}\left(\alpha \right)}$ in the estimate \eqref{Estim 1}.
Thus we conclude that as being additionally affected by two small factors, the maximal value of the second polarization  component in 180$^0$ walls in BTO is expected to be two orders of magnitude smaller than $P_s$. We would like to stress that there is no reason to expect anomalously small values of this component in ferroelectrics in general.

\section{Numerical Results as applied to barium titanate}
\label{sec:Generalized}

 To be more specific, we analyze the impact of the flexoelectric coupling on the structure of a domain wall in ferroelectrics for the case of the tetragonal phase in classical perovskite ferroelectric BaTiO$_3$ at room temperature.
 Since the problem is not analytically tractable, we do it numerically using the thermodynamic parameters of this crystal.
 Since the experimental values of components of the flexocoupling tensor $F_{ijkl}$ are not currently available,  we use in our calculations the $F_{ijkl}$ tensor evaluated from the data of ab initio calculations for $\textrm{Ba}_{0.5}\textrm{Sr}_{0.5}\textrm{Ti0}_3$ (BST) from Ref. \cite{Ponomareva}.
 In case where the results are found critical to the exact value of this tensor, we vary these to cover possible situations.
 The parameters used for the simulations are listed in the Table 1.

\begin{table}[1]
\caption{Free energy coefficients for bulk ferroelectric BaTiO$_3$ (from Refs.\cite{Bell}, \cite{MartonRychetskyHlinka},\cite{Pertsev98}, \cite{Ponomareva}).}
\label{BTO Parameters}
\begin{center}
\begin{tabular}{|p{1.5in}|p{3.0in}|p{1.5in}|} \hline
parameters & values & Refs. and Notes \\ \hline
$a_1 (C^{-2}·mJ)$ & $a_1=3.34(T-381)\times 10^5$ & Ref. \cite{Bell}  \\ \hline
$a_{ij} \, (C^{-4}·m^5J)$ & $a_{11}\simeq4.69(T-393)\times 10^6-2.02\times10^8$ \newline $a_{12}\simeq 3.230 \times 10^8$ & Ref. \cite{Bell} \\ \hline
$a_{ijk}\, (C^{-6}·m^9J)$\textit{} & $a_{111}\simeq-5.52(T-393)\times 10^7+2.76\times10^9$ \newline $a_{112}=4.47\times10^9$\newline $a_{123}=4.91\times10^9 $& Ref. \cite{Bell}\textit{} \\ \hline
${D_{ij}^{(u)}} (C^{-2}m^3J)$  & $D_{11}^{(u)}=5.1\times10^{-10}$\newline $D_{12}^{(u)}\simeq-0.2\times10^{-10}$\newline $D_{66}^{(u)}\simeq0.2\times10^{-10}$ & Ref. \cite{MartonRychetskyHlinka} \\ \hline
$Q_{ij} (C^{-2}·m^4)$ & $Q_{11}=0.11, Q_{12}\simeq-0.043, Q_{66}=0.059$ & Ref. \cite{Pertsev98} \\ \hline
$s_{ij} (10^{-12} Pa^{-1})$ & $s_{11}=8.3, s_{12}\simeq-2.7, s_{66}=9.24$ & Ref. \cite{Pertsev98}  \\ \hline
$f_{ijkl}^{(BST)} (V) $ & $f_{11}=5.12, f_{12}=3.32, f_{66}=0.045 $ & Ref. \cite{Ponomareva}  \\ \hline
$F_{ijkl} (10^{-11}C^{-1}m^3) $ & $F_{11}=2.46, F_{12}=0.48, F_{66}=0.05 $ \newline $F_a^{(BST)} = F_{66} -F_{11} +F_{12}\approx -1.93$ \newline & Recalculated  using relationship $ F_{\alpha\gamma}=f_{\beta\gamma} s_{\beta\gamma}$; $ f_{\alpha\gamma}$ taken from Ref. \cite{Ponomareva}  \\ \hline
$D_{ijkl}^{\left(\sigma \right)}\equiv D_{ijkl}$ \newline $(C^{-2}m^3J)$ & $D_{11}^{\left(\sigma \right)}\equiv D_{11} =3.52\times10^{-10}$ \newline $D_{12}^{\left(\sigma \right)}\equiv D_{12} =-1.24 \times 10^{-10}$ \newline $D_{66}^{\left(\sigma \right)}\equiv D_{66} =0.2 \times 10^{-10}$ & recalculated using $ D_{\alpha\gamma}^{(\sigma)}=D_{\alpha\gamma}^{(u)}-f_{\alpha\beta} F_{\beta\gamma}$ \\ \hline
\end{tabular}
\end{center}
\end{table}

\begin{figure}
    \includegraphics[width=0.5\textwidth]{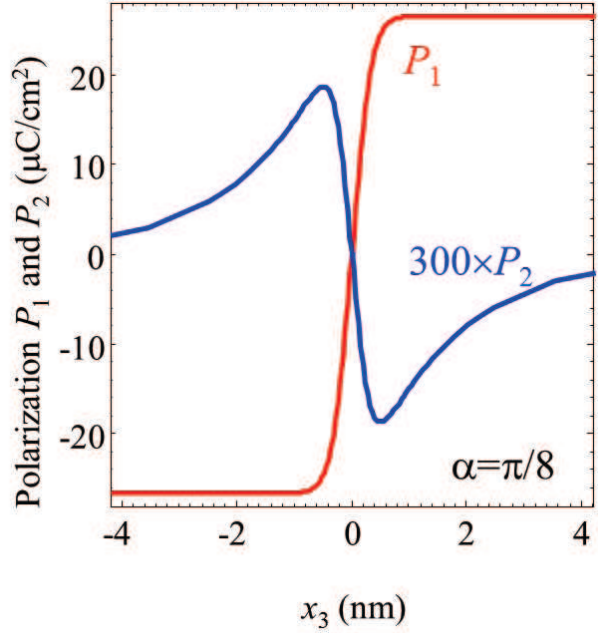}
    \caption{$P_1(x_3)$ and $P_2(x_3)$ - profiles across $180^o$  domain wall calculated for the parameters of BaTiO$_3$ at room temperature given in Table 1.  The $P_1(x_3)$ -profile is practically independent of the angle $\alpha$ between the wall and a cubic crystallographic direction. The $P_2(x_3)$ - profile is calculated for $\alpha = \pi/8$}.
    \label{P12 Profile}
\end{figure}

The profiles for $P_1$ and $P_2$ polarization components obtained from a numerical solution to Eqs. \eqref{Euler Alp} and \eqref{Sigma Through P} are shown in Fig. \ref{P12 Profile}.
We use \eqref{Pol Bound Cond} as the boundary conditions with an additional condition of vanishing of spatial derivatives of all variables at the infinity.
The $P_1$ profile is perfectly described by formula \eqref{P1 Approxim}, the difference conditioned by the flexoelectric effect is within the line width of the plot for any angle $\alpha$.
The maximal value of $P_2$ is as small as $6\cdot10^{-2} \mu C/cm^2$ which is $\sim P_S/300$. One can see from the Fig. \ref{P12 Profile} that the width of the domain wall with respect to $P_2$ is few times larger than with respect to $P_1$. This is a consequence of the small ratio of correlation lengths $\frac{r(P_1)}{r(P_2)} = \frac{\sqrt{D_{66}}}{\sqrt{D_{44}\left(\alpha \right)}}$ which results from $D_{11}/D_{66}\simeq 25$.
In the limit $P_2\ll P_S$ and $r(P_1) \ll r(P_2)$  an approximate analytical solution for the wall profile can be also developed, which is given in Appendix \ref{Analitica}.
\begin{figure}
    \includegraphics[width=0.9\textwidth]{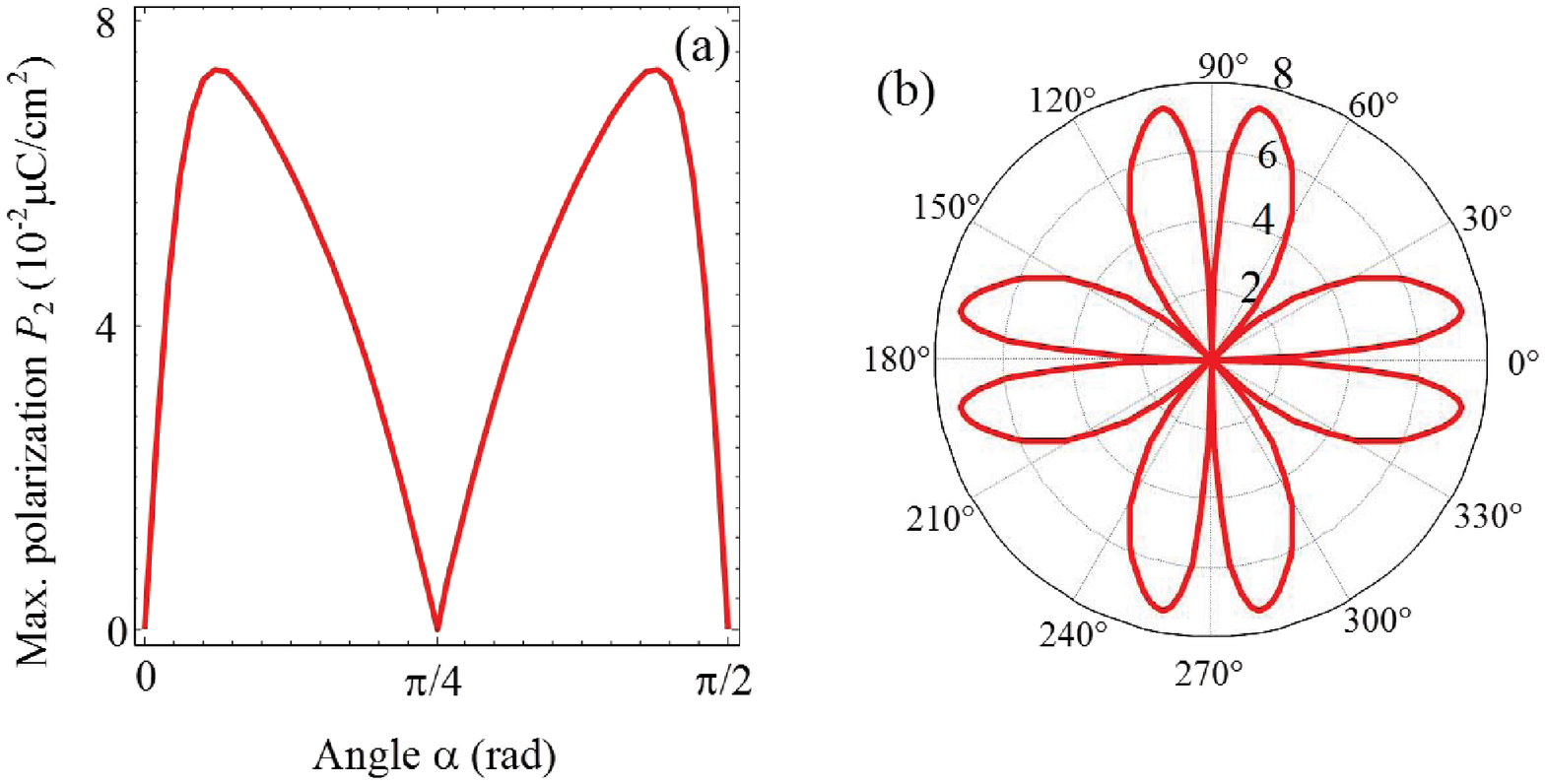}
    \caption{Maximal absolute value of of the second polarization component in the wall as a function of  the angle $\alpha$ between it and a cubic crystallographic direction calculated for the parameters of BaTiO$_3$ at room temperature given in Table 1. (a) - plot in one quadrant; (b) - polar plot.}
    \label{P2 Angle Dep}
\end{figure}
 Dependence of the maximal polarization $P_{2} \left(x_{3} \right)$ on the angle $\alpha$ is shown in Fig. \ref{P2 Angle Dep}. The anisotropy may be understood from expression \eqref{Approxim Amp F} as an interplay between the angular dependence of the flexoelectric factor $F_{24} \left(\alpha \right)$ and that of the gradient-energy factor $D_{44} \left(\alpha \right)$.

\begin{figure}
    \includegraphics[width=0.9\textwidth]{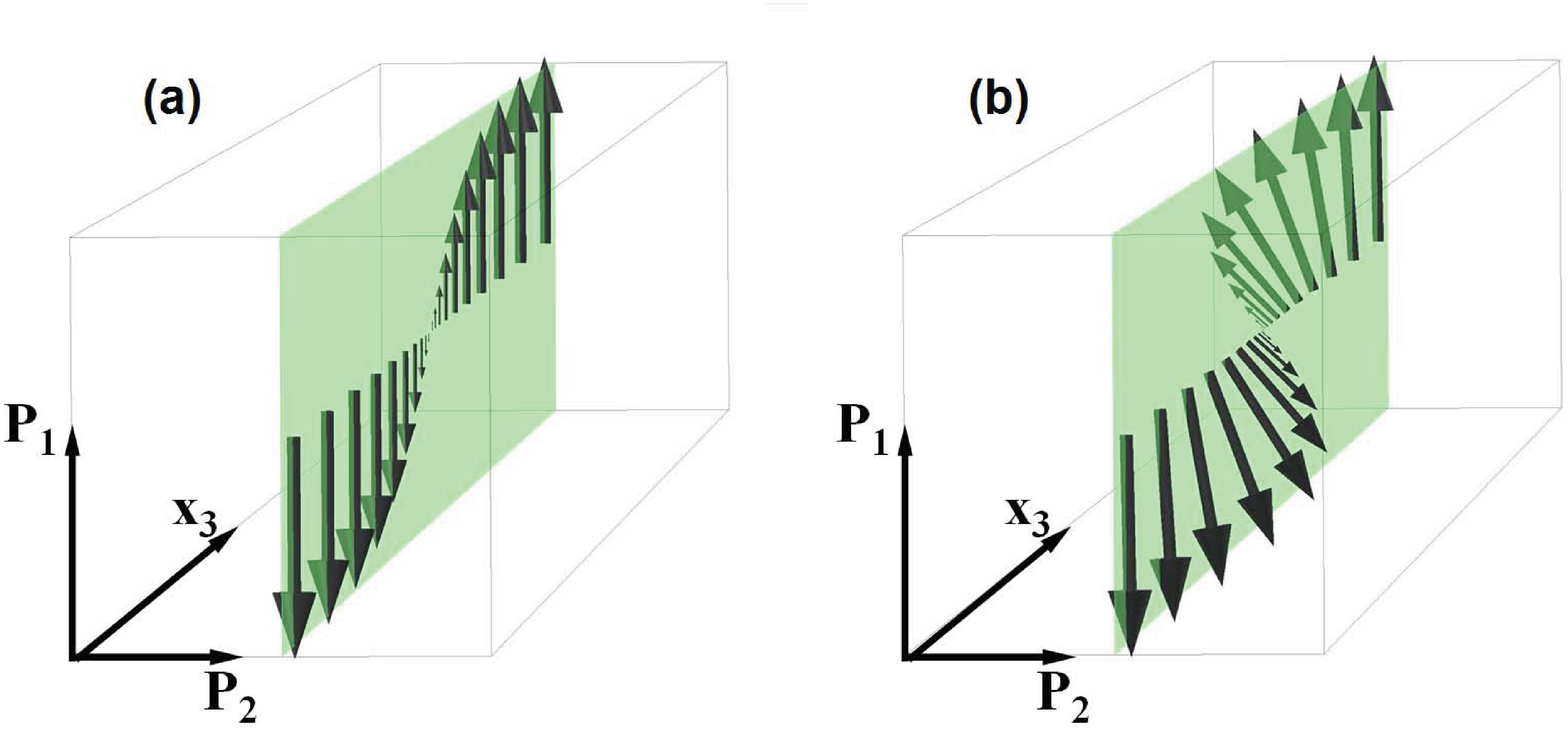}
    \caption{Schematic of the structures of  neutral 180-degree domain walls addressed in the paper. (a) - the Ising type structure occurring when the wall is normal to the cubic crystallographic directions or/and when the flexoelectric coupling is isotropic or neglected; (b) - the bichiral structure occurring for the oblique orientation of the wall provided that the flexoelectric coupling is anisotropic.}
    \label{3D Polarization profile}
\end{figure}
Inspection of the results of the simulations reveals that the flexoelectric coupling can lead to the formation of a polarization profile which has never been obtained to ferroelectric domain walls.
The polarization rotates in the boundary, but the latter does not acquire chirality.
The senses of the polarization rotation on the two sides of the boundary are opposite, the polarization passing though zero at its center.
As the result the polarization profile of the modified wall remains invariant with respect to the inversion about the wall center, in contrast to the Bloch-type wall.
Following the terminology introduced by Houchmandzadeh et al \cite{Houchmandzadeh1991} such wall can be classified as \textit{bichiral}.
One should note that the rough approximate solution \eqref{Approxim Sol 1} qualitatively correctly reproduced the bichirality of the wall.

\subsection{Free energy calculations }
\label{Free Enargy Calcul}

In this subsection we study the anisotropy of DW energy induced by flexoelectric effect.
The expression for the Free energy density $ \Phi$ may be obtained from Gibbs potential $G$ by Legendre transformation $\Phi=G+\sigma_i\varepsilon_i$.
The wall energy per unit area $E_W$ is then given by the integral $E_W= \int _{-\infty }^{+\infty }(\Phi(x_3)-\Phi_{\infty})dx_3$ where $\Phi_{\infty}$ is the energy density at $x_3\to\pm\infty$.
From \eqref{Gibbs_Alp} with the aide of \eqref{Spont Strains} and  \eqref{Sigma Through P} we derive the following expression for the wall free energy:

\begin{eqnarray}
\label{Wall Energy}
 E_W=\int _{-\infty}^{\infty}\{a_{1} \left(P_{1}^{2} -P_{S}^{2}+P_{2}^{2} \right)+a_{11} \left(P_{1}^{4}-P_{S}^{4} \right)+a_{22} \left(\alpha \right)P_{2}^{4}+a_{12} P_{1}^{2} P_{2}^{2}+\\ \nonumber +a_{111} \left(P_{1}^{6} -P_{S}^{6}\right)+a_{222} \left(\alpha \right)P_{2}^{6} +a_{112} P_{1}^{4} P_{2}^{2} +a_{122} \left(\alpha \right)P_{2}^{4}P_{1}^{2} + \\ \nonumber +\frac{D_{66} }{2} \left(\frac{\partial P_{1} }{\partial x_{3}^{} } \right)^{2} +\frac{1}{2} \left(D_{44} \left(\alpha \right)+\frac{F_{24}^{2} \left(\alpha \right)s_{11} }{s_{22} \left(\alpha \right)s_{11} -s_{12}^{2} } \right)\left(\frac{\partial P_{2} }{\partial x_{3} } \right)^{2} + \\ \nonumber +F_{24} \left(\alpha \right)\frac{Q_{12} s_{11} -Q_{11} s_{12} }{s_{22} \left(\alpha \right)s_{11} -s_{12}^{2} } \left(P_{S}^{2} -P_{1}^{2} \right)\frac{\partial P_{2} }{\partial x_{3} } + \frac{\left(Q_{11} \left(P_{S}^{2} -P_{1}^{2} \right)-Q_{12} P_{2}^{2} \right)^{2} }{2s_{11} } +\\ \nonumber+  \frac{\left(\left(Q_{12} s_{11} -Q_{11} s_{12} \right)\left(P_{S}^{2} -P_{1}^{2} \right)+P_{2}^{2} \left(-s_{11} Q_{22} \left(\alpha \right)+Q_{12} s_{12} \right)\right)^{2} }{2s_{11} \left(s_{22} \left(\alpha \right)s_{11} -s_{12}^{2} \right)}+ \frac{Q_{66}^{2} }{2s_{66} } P_{1}^{2} P_{2}^{2}\}\cdot dx_3
 \end{eqnarray}

\begin{figure}
    \includegraphics[width=0.5\textwidth]{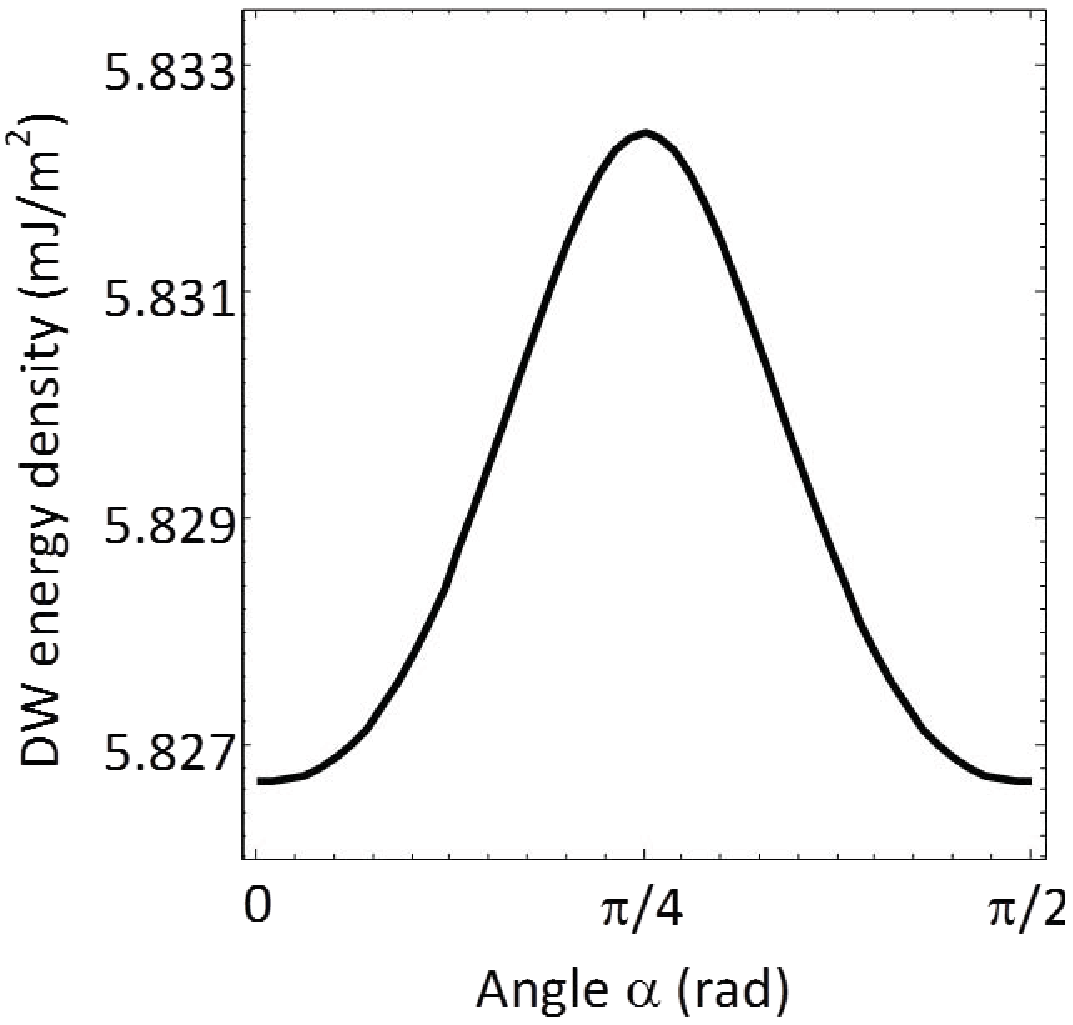}
    \caption{Energy of bichiral wall as a function of  the angle $\alpha$ between it and a cubic crystallographic direction.  Parameters used in the calculations are listed in Table 1 at room temperature T=293~K.}
    \label{EnergyAniso}
\end{figure}
\begin{figure}
    \includegraphics[width=0.5\textwidth]{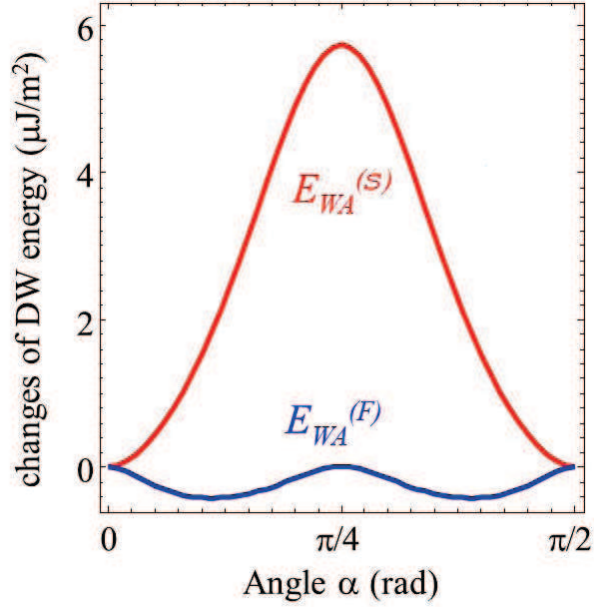}
    \caption{Comparison of the anisotropic contributions to the energy of bichiral wall as functions of  the angle $\alpha$ between it and a cubic crystallographic direction.
    $E_{WA}^{(s)}$ is the contribution solely controlled by the elastic anisotropy. $E_{WA}^{(F)}$ is the contribution due to the anisotropy of the flexoelectric coupling.
    The calculations are done for the parameters listed in Table 1 at room temperature T=293 K.}
    \label{EnergyContributions}
\end{figure}
\begin{figure}
    \includegraphics[width=0.5\textwidth]{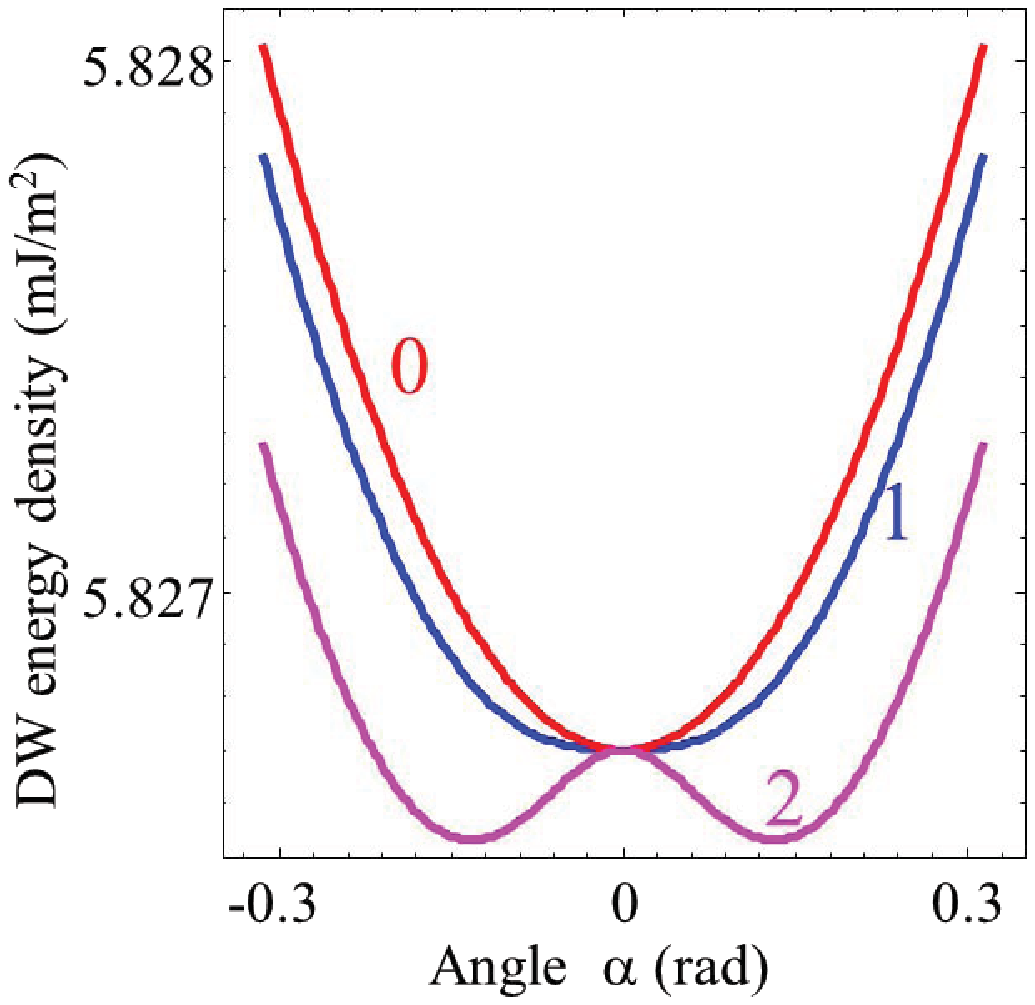}
    \caption{Energy of bichiral domain wall as a function of its orientation for different values  of the flexoelectric coupling calculated for the parameters of BTO (Table 1) at room temperature T=293.
    The curves labeled with 0, 1, and 2 correspond to the values of the anisotropic part of the flexoelectric coupling   $F_{a}$ equal to $0$, $F_{a}^{(BST)}$, and $2F_{a}^{(BST)}$ , where $F_{a}^{(BST)}$ is the value of the flexoelectric coefficient for BST recalculated using the results of ab initio from Ref. \cite{Ponomareva} (see Table 1).}
    \label{Splitting}
\end{figure}

Using the coordinate dependences obtained numerically and this relationship, we calculate the energy of the bichiral wall as a function of  the angle $\alpha$ plotted in the Fig. \ref{EnergyAniso}.
It is seen from this figure that the anisotropy of the wall energy is extremely weak.
However, it is worth looking closer the fine stricture of the anisotropic part of the wall energy, specifically to decompose it to the parts independent of and dependent on the flexoelectric coupling $E_{WA}^{(s)}$ and $E_{WA}^{(F)}$, respectively.
Such decomposition is presented in Fig.\ref{EnergyContributions} for the case where for the  flexoelectric coefficients of BTO we use those obtained by ab initio calculations for BST.
As we can see there is an essential difference between the $E_{WA}^{(s)}$ and $E_{WA}^{(F)}$ anglular dependences.  $E_{WA}^{(s)}$ is minimal at $\alpha=0$, maximal at $\alpha=\pi/4$ and monotonic in the interval $(0,\pi/4)$, while $E_{WA}^{(F)}$ is maximal in both  $\alpha=0$ and $\alpha=\pi/4$, and has a minimum in an intermediate point, around $\alpha=\pi/10$.
Remarkably, for small angles $\alpha$ the both contributions are comparable, the flexoelectric one being a bit smaller.
However if we take the flexoelectric factor $F_a$ two times as large as that for BST, flexoelectric contribution becomes dominating at small $\alpha$.
This leads to qualitative change of the anisotropy of the wall energy: flexoelectric coupling results in splitting  each of the energy minima (at $\alpha=0$ and $\alpha=\pi/2$) into two, see Fig. \ref{Splitting}.

The results of the numerical calculations can be elucidated using some analytical relationships.
First, the contribution $E_{WA}^{(s)}$ can be readily presented in a simple form
 \begin{eqnarray}
\label{Wall Energy s}
E_{WA}^{(s)} = \frac{Q_{11}^2}{2s_{11}}\left(\frac{Q_{12}}{Q_{11}}-\frac{s_{12}}{s_{11}}\right)^2 \frac{s_{11}^2 }{s_{22} \left(\alpha \right)s_{11} -s_{12}^{2} }\int _{-\infty}^{\infty} \left(P_s^2-P_1^2 \right)^2\cdot dx_3
 \end{eqnarray}
which is consistent with the results by  Dvorak and Janovec \cite{Dvorak}.
As was recognized by these authors (and clear from this expression), the angular dependence of this contribution is conditioned by that of the elastic compliance.
The flexoelectricity-conditioned contribution can be evaluated taking into account that in the case of BTO $P_2 \ll P_s$ and $r(P_1) \ll r(P_2)$. It enables us to keep among $P_2$-containing terms in \eqref{Wall Energy} only  $dP_2/dx_3$  (one can check that terms containing $P_2^2$ are smaller by factor $\frac{D_{66}}{D_{44}\left(\alpha \right)}$) while using \eqref{Euler2 Approxim 2} as an approximate relationship for $dP_2/dx_3$ to get:

 \begin{eqnarray}
\label{Wall Energy F}
 E_{WA}^{(F)}\approx E_{WA}^{(s)}\frac{2F_{24}^{2} \left(\alpha \right)s_{11}}{\tilde{D}_{44}\left(\alpha \right)(s_{22} \left(\alpha \right)s_{11} -s_{12}^{2})}.
 \end{eqnarray}

It is clear from this relationship that the smallness of both anisotropic contributions to the wall energy is controlled by the square of the same factor $Q_{12}/Q_{11}-s_{12}/s_{11}\,\, \simeq1/15$, which was already recognized responsible for the smallness of the second polarization component.
As for the shape of the angular dependence of the wall energy, it is conditioned by an interplay among the elastic anisotropy, anisotropy of the correlation energy, and that of the flexoelectric coupling.
\section{Conclusions}
The presented analysis demonstrates that the flexoelectric coupling once it is anisotropic (at $F_a=F_{11}-F_{12}-F_{66}=0$ all effects addressed in the paper disappear) results in the formation of 180-degree ferroelectric  domain walls  where the polarization rotates but the wall does not acquire chirality.
Following the terminology introduced by Houchmandzadeh et al \cite{Houchmandzadeh1991} such wall can be classified as bichiral.
In contact to the Bloch type walls, the appearance of the second polarization component in bichiral walls dose not brake the wall symmetry with respect to the  spatial inversion.
In addition, depending on the value of $F_a$, the flexoelectric coupling can lead to the doubling of the number of the energetically favorable orientations of the walls.
Order-of-magnitude estimates show that, in general,  the effects driven by the flexoelectric coupling can be appreciable.
However, for the thermodynamic parameters of barium titanate, the calculations performed show that the amplitude of the second component is expected to be smaller than one percent of the spontaneous polarization while the modulation of the wall energy (as a function of its orientation) is found to be yet smaller.
The numerical smallness of these effects is shown to be mainly conditioned by that of the $q_{12}$ component of the electrostriction "strain-polarization" tensor, which is typical for metal-oxide ferroelectric perovskites.
We would like to stress that there is no reason to expect the flexoelectricity-induced features of 180-degree ferroelectric  domain walls  addressed in this paper to be anomalously small in ferroelectrics in general.

\section{Acknowledgement}

E.A.E and A.N.M. are thankful to NAS Ukraine and NSF-DMR-0908718 for support.
P.V.Y., A.K.T and N.S. acknowledge Swiss National Science foundation for financial support.
\appendix
  \section{Approximate analytical solution for the wall profile }\label{Analitica}

Let us derive an analytical solution for the polarization profile in the approximation $P_2\ll P_S$; $r(P_1) \ll r(P_2)$. We solve Eq.\eqref{Euler2 Linear} taking $P_{1}^{2} \to P_{S}^{2} $ everywhere except the derivative $\frac{\partial \sigma _{2} }{\partial x_{3} } $, which is the driving force for the $P_2$ appearance. This approximation is valid far from the wall center where $P_1\approx P_s$, and in the wall center where the gradient term is dominating over the term distorted by $P_{1}^{2} \to P_{S}^{2} $.

The solution to the second order differential equation with constant coefficients, coordinate dependent inhomogeneity and boundary conditions \eqref{Pol Bound Cond} could be written using Green function method \cite{Korn}. Rewriting the equation \eqref{Euler2 Linear} for $P_2$ in the from

\begin{equation}
P_{2}^{} -r^{2} \frac{\partial ^{2} P_{2} }{\partial x_{3}^{2} } = \frac{F_{24} \left(\alpha \right)}{2a_{1} +\left(2a_{12} +\left({Q_{66}^{2}  \mathord{\left/{\vphantom{Q_{66}^{2}  s_{66} }}\right.\kern-\nulldelimiterspace} s_{66} } \right)\right)P_{S}^{2} +2a_{112} P_{S}^{4} } \frac{\left(Q_{12} s_{11} -Q_{11} s_{12} \right)}{s_{22} \left(\alpha \right)s_{11} -s_{12}^{2} } \frac{\partial \left(P_{S}^{2} -P_{1}^{2} \right)}{\partial x_{3} }
\end{equation}
where

\begin{equation}
\label{Approxim Sol 2 r}
r=\sqrt{\frac{\tilde{D}_{44} \left(\alpha \right)}{2a_{1} +\left(2a_{12} +\left({Q_{66}^{2}  \mathord{\left/{\vphantom{Q_{66}^{2}  s_{66} }}\right.\kern-\nulldelimiterspace} s_{66} } \right)\right)P_{S}^{2} +2a_{112} P_{S}^{4} } }
\end{equation}
one could readily\footnote{ The equation  $P-r^{2} \frac{d^{2} P}{dx^{2} } =E$  has a solution in the from  $P\left(x\right)=\frac{1}{2r} \int _{-\infty }^{+\infty }\exp \left(-\frac{\left|x-\xi \right|}{r} \right) E\left(\xi \right)d\xi $ . The only condition here is the absence of the solution at the infinity.} find that

\begin{eqnarray}
\label{P2 Prom}
P_{2} \left(x_{3} \right)  =  \frac{1}{2r} \int _{-\infty }^{+\infty }\exp \left(-\frac{\left|x_{3} -\xi \right|}{r} \right) \frac{\partial \left(P_{S}^{2} -P_{1}^{2} \left(\xi \right)\right)}{\partial \xi } d\xi \cdot  \\ \nonumber  \cdot \frac{F_{24} \left(\alpha \right)}{2a_{1} +\left(2a_{12} +\left({Q_{66}^{2}  \mathord{\left/{\vphantom{Q_{66}^{2}  s_{66} }}\right.\kern-\nulldelimiterspace} s_{66} } \right)\right)P_{S}^{2} +2a_{112} P_{S}^{4} } \frac{\left(Q_{12} s_{11} -Q_{11} s_{12} \right)}{s_{22} \left(\alpha \right)s_{11} -s_{12}^{2}}  \end{eqnarray}

Using an approximation for $P_1$ profile \eqref{P1 Approxim}

\begin{equation}
\label{P1 Approxim Approxim}
P_{1} \approx P_{S} \left(1-\exp \left({-\left|x_{3} \right| \mathord{\left/{\vphantom{-\left|x_{3} \right| R_{c} }}\right.\kern-\nulldelimiterspace} R_{c} } \right)\right)\textrm{sign}(x_{3} )
\end{equation}
we obtained from Eq.(15b) the following expression:
\begin{subequations}
\label{Approxim Sol 2}
\begin{eqnarray}
\label{Approxim Sol 2 Main}
P_{2} \left(x_{3} \right)\approx \frac{-f\left(\alpha \right)P_{S}^{2} p\left(x_{3} \right)}{2a_{1} +\left(2a_{12} +\left({Q_{66}^{2}  \mathord{\left/{\vphantom{Q_{66}^{2}  s_{66} }}\right.\kern-\nulldelimiterspace} s_{66} } \right)\right)P_{S}^{2} +2a_{112} P_{S}^{4} } \\                 \label{Approxim Sol 2 P}
p\left(x_{3} \right)=\frac{-2R_{c} }{R_{c}^{2} -r^{2} } \left(\exp \left(-\frac{x_{3} }{R_{c} } \right)-\exp \left(-\frac{x_{3} }{r} \right)\right)+\\ \nonumber +\frac{2R_{c} }{R_{c}^{2} -4r^{2} } \left(\exp \left(-\frac{2x_{3} }{R_{c} } \right)-\exp \left(-\frac{x_{3} }{r} \right)\right)
\end{eqnarray}
\end{subequations}
\begin{figure}
    \includegraphics[width=0.5\textwidth]{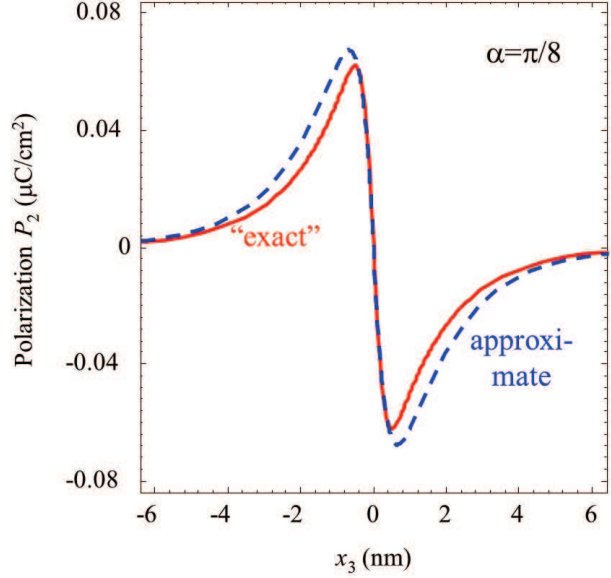}
    \caption{$P_{2} \left(x_{3} \right)$-profile near $180^o$ DW in BaTiO$_3$ calculated for angle $\alpha=\pi/8$  and parameters from the \textbf{Table 1 }and room temperature. Approximate analytical expression \eqref{Approxim Sol 2} (dashed curves) and numerical simulations (solid curve) are shown.}
    \label{P2 Profile}
\end{figure}

Comparison of the approximate analytical solution \eqref{Approxim Sol 2}  with numerical calculations based on the coupled equations \eqref{Euler Alp} is shown in Fig. \ref{P2 Profile}. The difference between the approximate analytical expression \eqref{Approxim Sol 2} (dashed curve) and numerical simulation (solid curve) is of the order of several percent. Thus, for $P_{2} \left(x_{3} \right)$ analytical expression \eqref{Approxim Sol 2} works with sufficient accuracy.

\bibliography{FlexoNeutralWall_v25}

\end{document}